\documentclass[final,5p,times,twocolumn]{elsarticle}

\usepackage{graphicx}
\usepackage{amsmath,amssymb}
\usepackage{braket}
\usepackage{hyperref} 
\usepackage{subfigure}
\usepackage{bm} 
\usepackage{mathtools} 
\usepackage{float}
\usepackage{cprotect}
\usepackage{placeins} 

\newcounter{bla}

\journal{Computer Physics Communications}

\def\Re{{\rm Re\,}}

\def\p{\partial}

\begin{document}

\begin{frontmatter}



\title{MiTMoJCo: Microscopic Tunneling Model for Josephson Contacts}


\author[a]{D. R. Gulevich}

\address[a]{ITMO University, St. Petersburg 197101, Russia}

\begin{abstract}
MiTMoJCo (Microscopic Tunneling Model for Josephson Contacts) is C code which aims to assist modeling of superconducting Josephson contacts based on the microscopic tunneling theory. 
The code offers implementation of a computationally demanding part of this calculation, that is evaluation of superconducting pair and quasiparticle tunnel currents from the given tunnel current amplitudes (TCAs) which characterize the junction material. MiTMoJCo comes with a library of pre-calculated TCAs for frequently used Nb-AlOx-Nb and Nb-AlN-NbN junctions, a Python module for developing custom TCAs, supplementary optimum filtration module for extraction of a constant component of a sinusoidal signal and examples of modeling few common cases of superconducting Josephson contacts.
\end{abstract}




\end{frontmatter}



{\bf PROGRAM SUMMARY}

\begin{small}
\noindent
{\em Program Title:}  MiTMoJCo: Microscopic Tunneling Model for Josephson Contacts                                   \\
{\em Licensing provisions(please choose one): } GPLv3                                  \\
{\em Programming language: } C, Python                                   \\
{\em Nature of problem:} 
	Modeling superconducting Josephson contacts based on the microscopic tunneling theory; calculation and fitting of the tunnel current amplitudes.
	\\
{\em Solution method:} 
  Computationally efficient Odintsov-Semenov-Zorin algorithm is used to account for the memory effects in pair and quasiparticle tunnel currents.
  \\
\end{small}

\newpage


\section{Introduction}

The microscopic tunneling theory (MTT) developed by Werthamer and others~\cite{Cohen, AmbBar, Werthamer, Larkin} is fundamental for description of the tunnel phenomena in superconducting Josephson contacts.
The theory played an essential role in development of SIS (superconductor-insulator-superconductor) mixers~\cite{Tucker-IEEE, Tucker-Feldman} which are now common front-end elements in mm and sub-mm receivers.
The theory of Tucker based on the MTT is now a standard tool for description of 
quasiparticle tunnel currents in SIS detector systems.
Despite this success, the present theoretical treatment of Josephson tunneling phenomena 
remains largely based on a simplistic resistively shunted junction (RSJ) model and perturbed sine-Gordon equation (PSGE),  that is generalization of RSJ to extended junctions. 
Indeed, while the RSJ model can be derived from the MTT under some very special conditions~\cite{Likharev}, its widespread use, however, is significantly much more broad than what can be justified by the MTT. 
While the use of RSJ model may be explained by its ability to provide a simple qualitative description of tunnel phenomena in the low-frequency regime~\cite{DRGulevich-PhysicaC}, it use in description of Josephson junction dynamics is rather based on the computational difficulties posed by the MTT 
than accuracy or completeness of the theoretical description which RSJ may not guarantee.
The situation is more severe in case of Josephson junctions of a finite spatial extent:
the need to evaluate tunnel currents at all points of the extended junction which depend on the previous evolution
makes the analysis significantly more complicated than solving local in time PSGE.



An efficient algorithm for computer modeling of Josephson tunnel junctions using the MTT
was proposed 
by Odinstov, Semenov and Zorin (OSZ)~\cite{OSZ-0, OSZ}. Unfortunately, 
the early attempts of using the algorithm had not been very successful and 
raised concerns 
about its applicability to realistic Josephson systems~\cite{Hattel-1993}.
The cause for disagreement between the theoretical and numerical results obtained by~\cite{Hattel-1993,GJ-1992} lies, however, not in the OSZ algorithm itself, but poorly fitted tunnel current amplitudes used as an input to the OSZ procedure.
MiTMoJCo code which I present here combines the computational efficiency of the OSZ algorithm with reliability of the MTT without compromising its accuracy in modeling realistic superconducting systems.
MiTMoJCo is available under the GNU General Public License v3.0 at GitHub~\cite{mitmojco}.





\section{Theory behind MiTMoJCo}


MiTMoJCo aids evaluation of the tunnel current density through a Josephson tunnel junction,
\begin{equation}
j(\mathbf{r},t) = \alpha_N \frac{\p\varphi}{\p t}  + \bar{j}(\mathbf{r},t)
\label{jfull}
\end{equation}
where
\begin{equation}
\begin{split}
\bar{j}(\mathbf{r},t)=
\frac{k}{\Re \tilde{j}_p(0) }
\int_0^{\infty}\Big\{ 
j_p(kt')\,\sin\left[\frac{\varphi(\mathbf{r},t)+\varphi(\mathbf{r},t-t')}{2}\right] \\
+\, \bar{j}_{qp}(kt')\,\sin
\left [ \frac{\varphi(\mathbf{r},t)-\varphi(\mathbf{r},t-t')}{2}\right ] 
\Big\}\; dt',
\end{split}
\label{MM2D-jbar}
\end{equation}
is the {\it reduced} tunnel current density which, by construction, depends on the history of evolution of the superconducting phase difference~$\varphi(\mathbf{r},t)$.
The tunnel current density~\eqref{jfull} 
enters the integro-differential equation describing dynamics of $\varphi(\mathbf{r},t)$ inside the Josephson junction,
\begin{equation}
\frac{\p^2\varphi}{\p t^2} -\left(1+\beta\frac{\p}{\p t} \right) \nabla^2 \varphi + \alpha_N \frac{\p\varphi}{\p t} + \bar j(\mathbf{r},t) =0
\label{MM2D}
\end{equation}
$$
\mathbf{n}\cdot\left(1+\beta\frac{\partial}{\partial t}\right)\nabla\varphi  = 
\mathbf{e}_z \cdot \left[\mathbf{n}\times\mathbf{h}  \right]
$$
where $\mathbf{n}$ is the in-plane outward normal, $\mathbf{h}$ is the normalized magnetic field in units~$j_c \lambda_J$.
The time-domain kernels $j_p(\tau)$ and $\bar{j}_{qp}(\tau)$
are defined by Fourier transforms of the {\it tunnel current amplitudes} $\tilde{j}_p(\xi)$ and $\tilde{j}_{qp}(\xi)$ which can be calculated theoretically from the Bardeen-Cooper-Schrieffer (BCS) theory and will discussed in more detail in Section~\ref{sec:tca}.
The parameter $k=\omega_g/\omega_J$ is the normalized gap frequency in units of the Josephson plasma frequency.
In Eqs.~\eqref{MM2D} and~\eqref{MM2D-jbar} time is measured in units of the inverse angular Josephson plasma frequency $\omega_J^{-1}$, the spatial coordinates are expressed in units of the Josephson penetration length~$\lambda_J$, the reduced current density
$\bar{j}(\mathbf{r},t)$ is normalized to $V_g/A R_N$, where $V_g$ is the gap voltage, $A$ is the total area of the junction and $R_N$ is the normal resistance.
The bar over the reduced tunnel current density $\bar{j}(\mathbf{r},t)$ and a reduced quasiparticle time-domain kernel $\bar{j}_{qp}(\tau)$ signifies that the normal resistance contribution has been subtracted: it enters to the full tunnel current~\eqref{jfull} and the equation~\eqref{MM2D} explicitly as a damping term $\alpha_N\partial\varphi/\partial t$. This is done for computational reasons to avoid the singularity at $\tau=0$ and obtain a regular behaviour of the quasiparticle time-domain kernel as a function of time. 
Furthermore, this allows constructing convenient semi-implicit numerical schemes where part of the tunnel current (the term $\alpha_N\partial\varphi/\partial t$) is computed implicitly.


While the numerical integration of the differential part of the integro-differential equation~\eqref{MM2D} is straightforward using the standard finite difference or finite element methods, its integral part poses significant computational difficulties due to the need to evaluate the memory integrals~\eqref{MM2D-jbar} at each time step of the numerical scheme. MiTMoJCo takes a full care of evaluating the memory part of this calculation by implementing the OSZ algorithm in C programming language.

MiTMoJCo enables to perform numerical calculations which are specific to given superconducting materials which constitute the junction. Such information is contained in the tunnel current amplitudes (TCAs) which are related to 
the time-domain kernels~\eqref{MM2D-jbar} by Fourier transforms
\begin{equation}
\begin{split}
\tilde{j}_{p}(\xi) & = \int_{-\infty}^\infty j_{p}(\tau) e^{-i \xi \tau}d\tau
\\
\tilde{j}_{qp}(\xi) & = i\xi + \int_{-\infty}^\infty \bar{j}_{qp}(\tau) e^{i \xi \tau}d\tau.
\label{transforms}
\end{split}
\end{equation}
Because $j_{p}(\tau)$ and $\bar{j}_{qp}(\tau)$ are real and satisfy causality conditions $j_{p}(\tau)=0$, $\bar{j}_{qp}(\tau)=0$ for $\tau<0$, their Fourier transforms are subjected to $\tilde{j}_{p}(-\xi) = \tilde{j}_{p}(\xi)^*$ and Kramers-Kronig-like dispersion relations~\cite{Zorin}.
Expressions for the TCAs for finite temperature $T\ge 0$ were calculated by Larkin and Ovchinnikov~\cite{Larkin} and are summarized in Ref.~\cite{DRGulevich-PRB-2017} (I refer to the verified expressions in Ref.~\cite{DRGulevich-PRB-2017} as the original ones in Ref.~\cite{Larkin} were given with a misprint). 
Bare BCS TCAs differ from experimental results for Nb-based junctions by~\cite{Likharev}:
(i) presence of sharp logarithmic singularities (Riedel peaks) and 
(ii) higher critical current values than observed in the experiments.
To remove these differences, a phenomenological procedure is applied which consists of smoothing 
the Riedel peaks~\cite{Zorin} and rescaling the pair current density~\cite{Zorin-1983}.


To use the power of the OSZ algorithms which considerably speeds up calculation of the memory integrals~\eqref{MM2D-jbar}, TCAs need to be fed to MiTMoJCo in the form of a sum of complex exponentials,
\begin{equation}
\begin{split}
j_{p}(\tau)=\Re \sum_{n=0}^{N-1} A_n\, e^{p_n \tau}
\\
\bar{j}_{qp}(\tau)=\Re \sum_{n=0}^{N-1} B_n\, e^{p_n \tau}
\end{split}
\label{exps}
\end{equation}
where $A_n$, $B_n$ and $p_n$ ($\Re p_n < 0$) are complex parameters. 
Values of the parameters can be obtained by fitting TCAs $\tilde{j}_{p}(\xi)$ and $\tilde{j}_{qp}(\xi)-i\xi$ by the Fourier transforms of the sums~\eqref{exps}.
It is possible to achieve a given precision of the fits by increasing the number of terms in~\eqref{exps}.
Typically, $N\sim 8$ is enough to obtain fits describing the true TCAs reasonably well, far beyond the precision at which TCAs derived from the BCS can be relied upon in description of realistic systems.




\section{MiTMoJCo C library interface}
\label{sec:interface}

MiTMoJCo offers object-oriented framework implemented in C.
MiTMoJCo functions are listed in~\ref{app-list} and are callable from both from C and C++ code.
At the heart of MiTMoJCo is the tunnel current object defined as a pointer \verb|TunnelCurrentType*|.
The type \verb|TunnelCurrentType| is a structure defined in the header~\verb|mitmojco.h|, 
\begin{verbatim}
    typedef struct {
        const char *filename;
        double a_supp;
        double kgap;
        double dt;
        int Ntotal;
        double *phi;
        int Nskip;
        int *skipinds;
        int Nnodes;
        int Nexps;
        MemState memstate;
        double Rejptilde0;
        double alphaN;
        double *jbar;
        void *self;
        bool error;
    } TunnelCurrentType;
\end{verbatim}

which has member variables:

\medskip
\verb|const char *filename|: pointer to tunnel current amplitude (TCA) file name (more on the tunnel current amplitudes in sec.~\ref{sec:tca}).

\verb|double a_supp|: pair current suppression parameter.

\verb|double kgap|: normalized gap frequency.

\verb|double dt|: integration time step.

\verb|int Ntotal|: size of the array phi.

\verb|double *phi|: pointer to the superconducting phase difference.

\verb|int Nskip|: number of shadow nodes to skip.
Often in numerical schemes one introduces shadow nodes used for treating the boundary conditions. However, the tunnel current only needs to be evaluated for the {\it physical nodes} and, therefore, its evaluation at the shadow nodes can be skipped. 

\verb|int *skipinds|: pointer to the array of node indices to skip.

\verb|int Nnodes|: number of active nodes, Nnodes = Ntotal-Nskip.

\verb|int Nexps|: number of exponentials used in fitting.

\verb|MemState memstate|: struct containing previous evolution information.

\verb|double Rejptilde0|: normalized critical current $\Re \tilde{j}_p(0)$.

\verb|double alphaN|: damping due to the normal resistance $\alpha_N$.

\verb|double *jbar|: pointer to the reduced current density array $\bar{j}(\mathbf{r},t)$. 


\verb|void *self| 
is used internally to access private member variables by the MiTMoJCo methods and is not intended for a regular user~\cite{self-footnote}. 

\verb|bool error|: handler to check that no errors occurred during the object creation (\verb|false| if no errors occurred an \verb|true| is error occurred such as e.g. missing amplitudes file or incorrect file format).

Pointer \verb|TunnelCurrentType*| is created by calling a constructor \verb|mitmojco_create|.
The constructor sets the first 8 elements of the structure \verb|TunnelCurrentType|:

\medskip

\begin{verbatim}
TunnelCurrentType* mitmojco_create (
    const char *filename, 
    double a_supp, 
    double kgap, 
    double dt, 
    int Ntotal, 
    double *phi, 
    int Nskip, 
    int *skipinds
    );
\end{verbatim}
As an example,
\begin{verbatim}
TunnelCurrentType *tunnel_current = 
    mitmojco_create( "NbNb_4K2_008.fit", 
    a_supp, kgap, dt, Nnodes, phi, 0, NULL );
\end{verbatim}
creates an object \verb|tunnel_current| based on the TCA file \verb|NbNb_4K2_008.fit|, suppression parameter given by the variable \verb|a_supp|, normalized gap frequency \verb|kgap|, time step of the numerical scheme \verb|dt|, number of spatial nodes \verb|Nnodes| and pointer \verb|phi| to an array with values of the superconducting phase difference at the nodes.
All member variables can be accessed via the arrow operator \verb|->|. For example, 
to access the pointer to values of the reduced tunnel current density $\bar{j}(\mathbf{r},t)$ use \verb|tunnel_current->jbar|.


\medskip

Upon creating the tunnel current object there are three methods available to the user:
\verb|mitmojco_init|, \verb|mitmojco_update| and \verb|mitmojco_free|.
Method \verb|mitmojco_init| initializes the state of the memory integrals assuming no dynamics in the past (that is, the supplied state is assumed to be stationary in which the system existed for an infinite time). Use
\begin{verbatim}
    mitmojco_init( tunnel_current );
\end{verbatim}
to initialize the tunnel current object based on the values of the \verb|phi| array whose address was supplied at the object construction call.
In the advanced case when dynamics in the past is given or known, the state of the memory integrals can be controlled via
the member structure \verb|memstate|.
Method \verb|mitmojco_update| updates values of the tunnel current at the physical nodes, based on the updated values of  \verb|phi| obtained within the numerical scheme. Call
\begin{verbatim}
    mitmojco_update( tunnel_current );
\end{verbatim}      
to update memory variables and values of the reduced tunnel current density $\bar{j}(\mathbf{r},t)$ which can be then be accessed via the object pointer as \verb|tunnel_current->jbar|.
Finally, \verb|mitmojco_free| is used to empty the memory allocated for the tunnel current object,
\begin{verbatim}
    mitmojco_free( tunnel_current );
\end{verbatim}

\medskip

Due to the object-oriented implementation, it is possible to deal with several tunnel current objects simultaneously, e.g. FFO and a SIS, two FFOs, an array of Josephson junctions etc. by creating independent objects each with its own parameters and different tunnel current amplitudes.

As in Josephson physics one often 
needs to calculate current-voltage characteristics, to assist its calculation MiTMoJCo offers a supplementary optimal filtration library which extracts efficiently the constant component of a periodic sinusoidal signal~\cite{OSZ}. The library is implemented in C and presented in~\ref{sec:opt_filter}.

\section{MiTMoJCo Python module for creation of tunnel current amplitudes}
\label{sec:mitmojco-module}

MiTMoJCo is provided with an easy-to-use module \verb|mitmojco| written in Python 3 for creation of custom fits of TCAs at arbitrary temperature, superconducting gaps of the materials and degree of Riedel peak smoothing.
The supplied Jupyter notebook \verb|amplitudes.ipynb| illustrates the use of \verb|mitmojco| module by creating a fit of TCAs.

The work with \verb|mitmojco| starts with the import statement
\begin{verbatim}
import mitmojco
\end{verbatim}
upon which the following functions of the \verb|mitmojco| module become available:
\begin{verbatim}
tca_bcs(T, Delta1, Delta2)
tca_smbcs(T, Delta1, Delta2, dsm)
new_fit(x, Jpair_data, Jqp_data, maxNterms, thr)
\end{verbatim}
Function \verb|tca_bcs| returns bare (without smoothing) BCS TCAs evaluated from the Larkin and Ovchinnikov expressions~\cite{Larkin} which are summarized in Ref.~\cite{DRGulevich-PRB-2017}; \verb|tca_smbcs| returns smoothed TCAs obtained by smoothing bare BCS TCAs using the smoothing procedure of Ref.~\cite{Zorin} (see also Ref.~\cite{DRGulevich-PRB-2017,DRGulevich-chaos-PRL,DRGulevich-NbN-FFO} for further details); \verb|new_fit| is used to calculate fits of exact TCAs and export them as a \verb|.fit| file in the format suitable for the use by MiTMoJCo C library.

\section{Pre-calculated tunnel current amplitudes}
\label{sec:tca}

Also, the user is supplied with pre-calculated fits of TCAs for Josephson junctions made by two common technologies, Nb-$\rm AlO_x$-Nb and Nb-AlN-NbN, at liquid helium boiling temperature 4.2 K, at difference values of the smoothing parameter. The pre-calcualted fits are summarized in Table~\ref{tab:amps}.
The tunnel current amplitudes are supplied without the account of the pair suppression which is controlled separately via parameter \verb|a_supp| from the MiTMoJCo interface (see section~\ref{sec:interface}).
The detailed information about precision of the fits can be found in the \verb|amplitudes| folder.

Some of the fits presented in Table~\ref{tab:amps} have been verified by numerical calculation of Josephson junction dynamics in several publications as summarized in the last column: fit~\verb|NbNb_4K2_008.fit| was used in Ref.~\cite{DRGulevich-PRB-2017},~\verb|NbNbN_4K2_008.fit| in Ref.~\cite{DRGulevich-chaos-PRL} and 
~\verb|NbNbN_4K2_015.fit| in Ref.~\cite{DRGulevich-NbN-FFO}.
For convenience, fits given in the Refs.~\cite{OSZ} and ~\cite{GJ-1992} are also provided.
Because of their poor performance in the subgap region these should not be used for production unless reproducing the results of Refs.~\cite{OSZ,GJ-1992,Hattel-1993} is your direct purpose.






\renewcommand{\arraystretch}{1.5}
\setlength{\tabcolsep}{10pt}
\begin{table*}[h!]
\begin{center}
    \begin{tabular}{| c | c | c | c | c | c | c | c | c |}
    \hline
    Filename & $T(\rm{K})$ & $\Delta_1(\rm{meV})$ & $\Delta_2(\rm{meV})$ & $\delta$ & $N$ & $\tau_{r}$ & $\tau_{a}$ & Source  \\ \hline
    \verb|NbNb_4K2_001.fit| & 4.2 & 1.40 & 1.40 & 0.001 & 10 & $0.005$ & $0.001$ & -- \\ \hline
    \verb|NbNb_4K2_002.fit| & 4.2 & 1.40 & 1.40 & 0.002 & 9 & $0.005$ & $0.001$ & -- \\ \hline
    \verb|NbNb_4K2_004.fit| & 4.2 & 1.40 & 1.40 & 0.004 & 9 & $0.004$ & $0.0008$ & -- \\ \hline
    \verb|NbNb_4K2_008.fit| & 4.2 & 1.40 & 1.40 & 0.008 & 8 & $0.005$ & $0.001$ & Ref.~\cite{DRGulevich-PRB-2017}\\ \hline    
    \verb|NbNb_4K2_016.fit| & 4.2 & 1.40 & 1.40 & 0.016 & 8 & $0.005$ & $0.001$ & -- \\ \hline
    \verb|NbNb_4K2_032.fit| & 4.2 & 1.40 & 1.40 & 0.032 & 8 & $0.004$ & $0.0008$ & -- \\ \hline
    \verb|NbNb_4K2_064.fit| & 4.2 & 1.40 & 1.40 & 0.064 & 8 & $0.005$ & $0.001$ & -- \\ \hline
    \verb|NbNbN_4K2_008.fit| & 4.2 & 1.40 & 2.30 & 0.008 & 8 & $0.010$ & $0.002$ & Ref.~\cite{DRGulevich-chaos-PRL}\\ \hline    
    \verb|NbNbN_4K2_015.fit| & 4.2 & 1.40 & 2.30 & 0.015 & 8 & $0.004$ & $0.0008$ & Ref.~\cite{DRGulevich-NbN-FFO} \\ \hline        
    \verb|OSZ_Table_1.fit| & -- & -- & -- & -- & 4 & -- & -- & Ref.~\cite{OSZ}  \\ \hline
    \verb|OSZ_Table_2.fit| & -- & -- & -- & -- & 5 & -- & -- & Ref.~\cite{OSZ}  \\ \hline
    \verb|GJHS_Table_1.fit| & 4.2 & 1.35 & 1.35 & -- & 4 & -- & -- & Ref.~\cite{GJ-1992}  \\ \hline
    \verb|GJHS_Table_2.fit| & 6.4 & 1.15 & 1.15 & -- & 5 & -- & -- & Ref.~\cite{GJ-1992} \\ \hline    
    \hline
    \end{tabular}
\end{center}    
    \caption{Library of pre-calculated fits of tunnel current amplitudes (TCAs) supplied with MiTMoJCo. Tunnel current amplitudes are calculated from the BCS 
    theory for Nb-$\rm AlO_x$-Nb and Nb-AlN-NbN junctions
and smoothed using different values of the phenomenological smoothing parameter $\delta$. Number of the fitting exponentials $N$, relative and absolute tolerances of the fit in the frequency region $|\xi|\le 2$ are also shown. 
I also provide fit files with parameters from the Tables 1 and 2 of Refs.~\cite{OSZ} and ~\cite{GJ-1992}. However, these should not be used for production purposes due to their inferior performance in the subgap region.
    } 
    \label{tab:amps}
\end{table*}







\section{Examples}
\label{sec:examples}

MiTMoJCo is provided with several examples of some common cases of modeling Josephson junctions.


\subsection{Example 1: Current-biased SIS junction}

Current-biased Josephson junction is described by
\begin{equation}
\ddot{\varphi} + \alpha_N \dot\varphi + \bar j(t) - \gamma = 0,
\label{cbJ}
\end{equation}
where $\gamma$ is the applied bias current and $\bar j(t)$ is given by~\eqref{MM2D-jbar} without the spatial dependence.
Using MiTMoJCo, the integro-differential equation~\eqref{cbJ} can be effectively treated 
by the standard finite difference schemes for an ordinary differential equation (ODE) where values of the integral $\bar j(t)$ at each time step are provided by MiTMoJCo. 
To access values of the tunnel current $\bar j(t)$, the object~\verb|sis_tunnel_current| should first be updated
at each time step,
\begin{verbatim}
    mitmojco_update( sis_tunnel_current );
\end{verbatim}
and then used to extract the up-to-date value
\begin{verbatim}
    sis_tunnel_current->jbar[0]
\end{verbatim}
of the tunnel current.
In the example 1 I use the 2nd order central difference discretization to solve~\eqref{cbJ}.
The resulting scheme turns out to be no different from the standard ODE in the presence of an independent driving force $\bar{j}(t)$, apart from the~\verb|mitmojco_update| statement which adjusts its value in accordance with the previous evolution.

 To compile the example, type
\begin{verbatim}
    $ make example-1
\end{verbatim}
which produces executable \verb|example-1| in the current directory.
Executing 
\begin{verbatim}
    $ ./example-1 1.1 0.0 0.01
\end{verbatim}
calculates the normalized dc voltage for a range of values of $\gamma$. In this case, $\gamma$ takes 1.1 as initial value (just above the critical current) and decreases down to 0.0 with step 0.01. At the execution, MiTMoJCo displays the details of the MTT model (details about the fitted TCAs, normalized gap frequency $k$, pair current suppression parameter $\alpha_{\rm supp}$, critical current $\Re\tilde{j}_p(0)$ and damping due to the normal resistance $\alpha_N$) and two columns corresponding to values of the current and voltage in normalized units.

\subsection{Example 2: Voltage-biased SIS junction under ac drive}

Current through a small voltage-biased Josephson junction is given by the Eq.~\eqref{jfull}. In this case, one does not need to solve the differential equation, rather, the superconducting phase difference $\varphi(t)$ can be easily found from the fundamental Josephson relation if dependence of the applied voltage on time is known.
In this case, all is needed in to update the tunnel current object~\verb|sis_tunnel_current| at every time step and 
add to the obtained $\bar{j}(t)$ the missing normal resistance part $\alpha_N\dot\varphi$ to restore the full tunnel current.

\begin{figure}[tbh!]
\begin{center}
\includegraphics[width=3.4in]{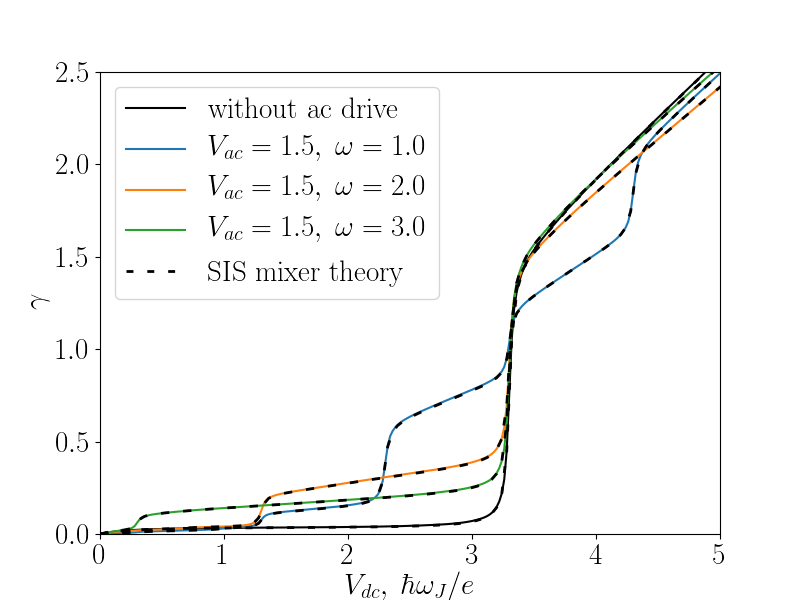}
\cprotect\caption{\label{fig:ex-2} 
IVC of voltage-biased SIS junction under ac drive calculated by MiTMoJCo in presence and absence of the ac drive. Parameters of the calculation are: tunnel current amplitudes file \verb|NbNb_4K2_008.fit|, pair current suppression $\alpha_{\rm supp}=0.7$, normalized gap frequency $k=3.3$. For verification of the MiTMoJCo calculation the known theoretical result from the SIS mixed theory~\cite{Tucker-Feldman} is also shown.
}
\end{center}
\end{figure}

In example 2 I assume the junction is driven by time-dependent voltage
$$
V(t) = V_{dc} + V_{ac} \cos(\omega t)
$$
To improve efficiency of determination of the constant voltage component $V_{dc}$
I recommend to use the optimum filtration routine for a sinusoidal signal~\cite{OSZ} offered by MiTMoJCo.
The resulting SIS IVC obtained at different values of the driving frequency $\omega$ is shown in Fig.~\ref{fig:ex-2}. 
In the simplest case of a harmonic drive discussed here, 
MiTMoJCo results coincide with those given by the SIS mixer theory where the explicit expressions in terms of the Bessel functions are known~\cite{Tucker-Feldman}. 

\subsection{Example 3: Sine-Gordon breather in long Josephson junction}

Consider a 1D model of a long Josephson junction,
\begin{equation}
\varphi_{tt} -\left(1+\beta\frac{\p}{\p t} \right) \varphi_{xx} + \alpha_N \varphi_t + \bar j(x,t) =0
\label{SG-breather}
\end{equation}
with open boundary conditions,
$$
\varphi_x(\pm L/2,t)=0.
$$
In analogy with the \verb|example-1| considered above, MiTMoJCo makes solution of integro-differential equation~\eqref{SG-breather} equivalent to solving partial differential equation (PDE) 
to which standard numerical apply. In this example I use a finite difference method obtained by 2nd order central difference discretization of spatial and time derivatives. Although, I use 1st order discretization for the surface damping term $\beta\varphi_{txx}$, the scheme order $O(\beta \Delta t)+O(\Delta t^2)$ remains effectively 2nd order as values of $\beta$ are typically smaller than the relevant values of the time step~$\Delta t$~\cite{DRGulevich-PRB-2017}.

\subsection{Example 4: Dynamics of a single fluxon}

Consider an annular Josephson of length $L\gg 1$ biased by a current $\gamma$. Similar to the previous case, the equation describing dynamics of the superconducting phase difference is
\begin{equation}
\varphi_{tt} -\varphi_{xx} + \alpha_N \varphi_t + \bar j(x,t) - \gamma =0
\label{SG-fluxon}
\end{equation}
$$
\varphi(L,t)=\varphi(0,t), \quad
\varphi_x(L,t)=\varphi_x(0,t)
$$
where I neglected the surface damping contribution.
The formalism of MTT has been applied to motion of an isolated fluxon in Refs.~\cite{GJ-1992,Hattel-1993}.
From power balance considerations Hattel et al.~\cite{Hattel-1993} derived the semi-analytical formula for fluxon velocity which 
suggest 
a higher slope of the current-velocity curve in the small velocity region.
In agreement with this, the current-velocity curve for a fluxon calculated by MiTMoJCo 
satisfies this theoretically predicted behavior.

\subsection{Example 5: Flux-Flow Oscillator}

Flux-flow oscillator~\cite{Nakajima-I} is a long Josephson junction where a moving chain of fluxons
is used to generate electromagnetic radiation.
Microscopic model of Nb-${\rm AlO_x}$-Nb and Nb-AlN-NbN flux-flow oscillators have been recently proposed and studied in Refs.~\cite{DRGulevich-PRB-2017,DRGulevich-NbN-FFO}. MiTMoJCo has been proven useful for calculation of current-voltage characteristics of realistic flux-flow oscillators in good agreement with experimental results~\cite{DRGulevich-PRB-2017,DRGulevich-NbN-FFO}.
The example 5 is based on the model of a flux flow oscillator~\cite{DRGulevich-PRB-2017} where coupling to a load has been removed for simplicity.
\section{deal.II+MiTMoJCo: 2D models of Josephson junction}

In combination with \verb|deal.II| the finite element library~\cite{dealii-original,dealii-8.5,dealii}
MiTMoJCo becomes a powerful tool for modeling of realistic 2D Josephson junctions.
The need to go beyond the 1D model arises in studies of transmission line intersections and networks~\cite{Nakajima-I,Nakajima-II,DGulevich-cloning,DGulevich-phenomena,DGulevich-NJP,DGulevich-pump,Caputo-cloning,Sobirov-2016}, 
curved waveguides~\cite{Gorria-2004,Olendski-2008}, radial soliton-like excitations~\cite{Bogolyubskii-1976,Bogolyubskii-1977,Chris-1978,Chris-1981,Chris-1997,Piette-1998},
superconducting chaotic oscillators~\cite{DRGulevich-chaos-PRL},
beam splitters~\cite{DGulevich-waveguides}, 2D fluxon oscillations modes~\cite{Lach-1993,DGulevich-shape-PRL,DGulevich-shape-PRB,Starodub-2014},
fluxon qubits~\cite{Wallraff-2000, Kemp-2002, Shaju-2004, Price-PRB-2010} and other systems.

MiTMoJCo contains an example of modeling T-junction Terahertz chaotic oscillator~\cite{DRGulevich-chaos-PRL} using the \verb|deal.II| library implemented in C++.

\section{Performance}

The serial version of the code is only about 2-3 times slower than the same numerical scheme implementing the standard time-local PSGE.
This is due to the fact that at each time step and spatial node MiTMoJCo evaluates two trigonometric functions~\cite{footnote} as opposed to one trigonometric function in PSGE, whereas updating the memory variables takes a relatively small percentage of CPU time.

The calculation of tunnel currents in MiTMoJCo is parallelized using OpenMP to run on the shared memory machines.
The parallel computation is implemented for junctions with a number of spatial nodes larger than 1. In this case evaluations of the tunnel currents at different spatial nodes are completely independent at each time step and allow for a straightforward parallel execution.

The performance of the presented \verb|deal.II|+MiTMoJCo model is fully limited by \verb|deal.II| functions implementing the finite element method where MiTMoJCo has a negligible effect on the CPU time. The tests has been done for the standard implementation of a time-dependent problem 
based on 
the matrix-vector products as described in Steps 23-25 of \verb|deal.II| examples. Having said this, 
the performance of the \verb|deal.II|+MiTMoJCo model have not been checked for the more advanced performance-tuned implementation based on the cell-based finite element quadrature proposed in Ref.~\cite{Martin-proc,Martin-2012} (see also Step 48 of \verb|deal.II|) which can influence favorably the performance of \verb|deal.II| in our model.

\section{Conclusion}



The presented code solves the problem of the high computational effort in modeling Josephson junctions using the MTT. MiTMoJCo makes numerical modeling based on the MTT 
not much different than solving ordinary or partial differential equations by the standard numerical methods.





\appendix

\section{MiTMoJCo optimum filtration library}
\label{sec:opt_filter}


Optimum filtration library for efficient calculation of a constant component of a sinusoidal signal is realized in C in the source \verb|opt_filter.c| and header \verb|opt_filter.h|. The code implements the algorithm outlined in Ref.~\cite{OSZ}. To access the optimum filtration routine 
begin with defining the filter object \verb|OptFilterType*| by calling its constructor \verb|opt_filter_create|.
The type \verb|OptFilterType| is defined in \verb|opt_filter.h| as a structure
\begin{verbatim}
    typedef struct {
        int n;
        double a;
        double *y;
        void *self;  
    } OptFilterType;
\end{verbatim}
The member variables \verb|n|, \verb|a| and the pointer to an array \verb|*y| correspond to $n$, $a$ and $y_n$ of Ref.~\cite{OSZ}. 
The constructor \verb|opt_filter_create| is called with a single argument which is the level of filtration given by $n$.
Parameter $n$ is integer and, in practice, takes values between 1 and 5, where 1 correspond to the arithmetic mean of the recorded values~\cite{OSZ}. The pointer \verb|void *self| points to a private structure not intended for ordinary user~\cite{self-footnote}.

The four methods accessible by the user are \verb|opt_filter_init|, \verb|opt_filter_update|, \verb|opt_filter_result| and 
\verb|opt_filter_free|.
Method \verb|opt_filter_init| initializes the filter object,
\begin{verbatim}
    opt_filter_init(voltage_filter);
\end{verbatim}
\verb|opt_filter_update| makes a record of the signal value,
\begin{verbatim}
    opt_filter_update(voltage_filter, voltage );
\end{verbatim}
\verb|opt_filter_result| returns value of the calculated dc component once the calculation is finished,
\begin{verbatim}
    Vdc = opt_filter_result(voltage_filter);
\end{verbatim}
and, finally, \verb|opt_filter_free| is used to clear the memory allocated to the filter object,
\begin{verbatim}
    opt_filter_free(voltage_filter);
\end{verbatim}

\section*{Acknowledgements}
The work is supported by the Russian Science Foundation under the grant 18-12-00429. The author is grateful to 
V. P. Koshelets for inspiring experiments which motivated development of MiTMoJCo, 
\verb|deal.II| online user group for the support during development of the \verb|deal.II|+MiTMoJCo model and
F. V. Kusmartsev for motivating discussions.


\begin{thebibliography}{99}
\bibitem{self-footnote} Presence of the \verb|void *self| pointer is a backside of the object-oriented implementation of a class in C as opposed to the more standard C++. Pointer \verb|self| points to a larger private structure \verb|TunnelCurrentType_Private| which contains both private variables used internally by MiTMoJCo as well as includes the interface to the public structure accessible by the user.
\bibitem{footnote} At each time step MiTMoJCo evaluates $\sin(\varphi_i/2)$ and $\cos(\varphi_i/2)$ for every spatial node $i$. In principle, this calculation can be further accelerated by evaluation of $\sin$ and $\cos$ via only one of the trigonometric functions.
\end{thebibliography}
\end{document}
